\documentclass{achemso}
 
\usepackage{graphicx}
\usepackage{float}

 \usepackage{subfigure}
 \usepackage{csquotes}
 \usepackage{amssymb}
 \usepackage{fixltx2e}

\newcommand{\onlinecite}[1]{\hspace{-1 ex} \nocite{#1}\citenum{#1}}

\author{Leonardo del Rosso}
\email{l.delrosso@ifac.cnr.it}
\affiliation{Consiglio Nazionale delle Ricerche, Istituto di Fisica Applicata  ``Nello Carrara'',\\via Madonna del Piano 10, I-50019 Sesto Fiorentino, Italy}

\author{Milva Celli}
\affiliation{Consiglio Nazionale delle Ricerche, Istituto di Fisica Applicata ``Nello Carrara'',\\via Madonna del Piano 10, I-50019 Sesto Fiorentino, Italy}

\author{Francesco Grazzi}
\affiliation{Consiglio Nazionale delle Ricerche, Istituto di Fisica Applicata  ``Nello Carrara'',\\via Madonna del Piano 10, I-50019 Sesto Fiorentino, Italy}

\author{Michele Catti}
\affiliation{Dipartimento di Scienza dei Materiali, Universit\`a di Milano Bicocca\\via Roberto Cozzi 55, I-20125 Milano, Italy}

\author{Thomas C. Hansen}
\affiliation{Institut Laue-Langevin (ILL), 71 avenue des Martyrs, 38000 Grenoble, France}

\author{Andrew Dominic Fortes}
\affiliation{ISIS Neutron and Muon Facility, Rutherford Appleton Laboratory, Harwell Science and Innovation Campus, Chilton, Oxfordshire OX11 OQX, U.K.}

\author{Lorenzo Ulivi}
\email{l.ulivi@ifac.cnr.it}
\affiliation{Consiglio Nazionale delle Ricerche, Istituto di Fisica Applicata  ``Nello Carrara'',\\via Madonna del Piano 10, I-50019 Sesto Fiorentino, Italy}

\title{Cubic ice Ic free from stacking defects synthesized from ice XVII}

\begin{document}


\begin{center}
\date{\today}
\end{center}

\normalsize

\newpage

\textbf{
Among the over eighteen different forms of water ice, only the common hexagonal phase and a cubic phase are present in nature on Earth.\cite{Hobbs74,Petrenko99}
The existence of these two polytypes, almost degenerate in energy, represents one of the most important and unresolved topics in the physics of ice.\cite{Salzmann11,Salzmann19,Bartels-Rausch12}
It is now widely recognised  that all the samples of ``cubic ice'' obtained so far are instead a stacking-disordered form of ice I (i.e. ice Isd), in which both hexagonal and cubic stacking sequences of hydrogen-bonded water molecules are present.\cite{Kuhs12,Malkin12,Malkin15}
Here we describe a new method to obtain cubic ice Ic in large quantities, and demonstrate its unprecedented structural purity from two independent neutron diffraction experiments performed on two of the leading neutron diffraction instruments in Europe.
}

Stacking disordered forms of cubic ice are generally prepared  by low-pressure vapour deposition\cite{Konig43}, or more commonly, by the back-transformation, at room pressure and low temperature, of amorphous\cite{Dowell60} or crystalline high-pressure ice polymorphs.\cite{Bertie63,Bertie64,Arnold68} 
We have prepared for the first time structurally pure ice Ic  by the transformation of a powder of ice XVII at room pressure by increasing temperature.
Ice XVII is a novel metastable phase of pure ice, obtained from the high-pressure hydrogen filled ice in the C$_0$-phase.\cite{del_Rosso16,del_Rosso16jpcc}
This low density solid water phase has the characteristic of being highly porous, and, unique among the various stable and metastable phases of ice, exhibits a structure comprising only pentagonal rings of water molecules.\cite{del_Rosso16jpcc,del_Rosso17prm}.
Ice XVII can be maintained at room pressure only up to about 130 K, above which it undergoes a phase transition similar to that mentioned above for the amorphous\cite{Dowell60} and high-pressure crystalline\cite{Bertie63,Bertie64} forms.
Whilst the end-product of all of these transitions, above 200 K, is the ordinary hexagonal form of ice (ice Ih), the remarkable difference between ice XVII and the other forms is the nature of the intermediate state, where, instead of stacking-disordered ice, we find a structurally-pure form of cubic ice (true ice Ic).

\begin{figure}[h]
\centering
 \includegraphics[width= 17.0cm]{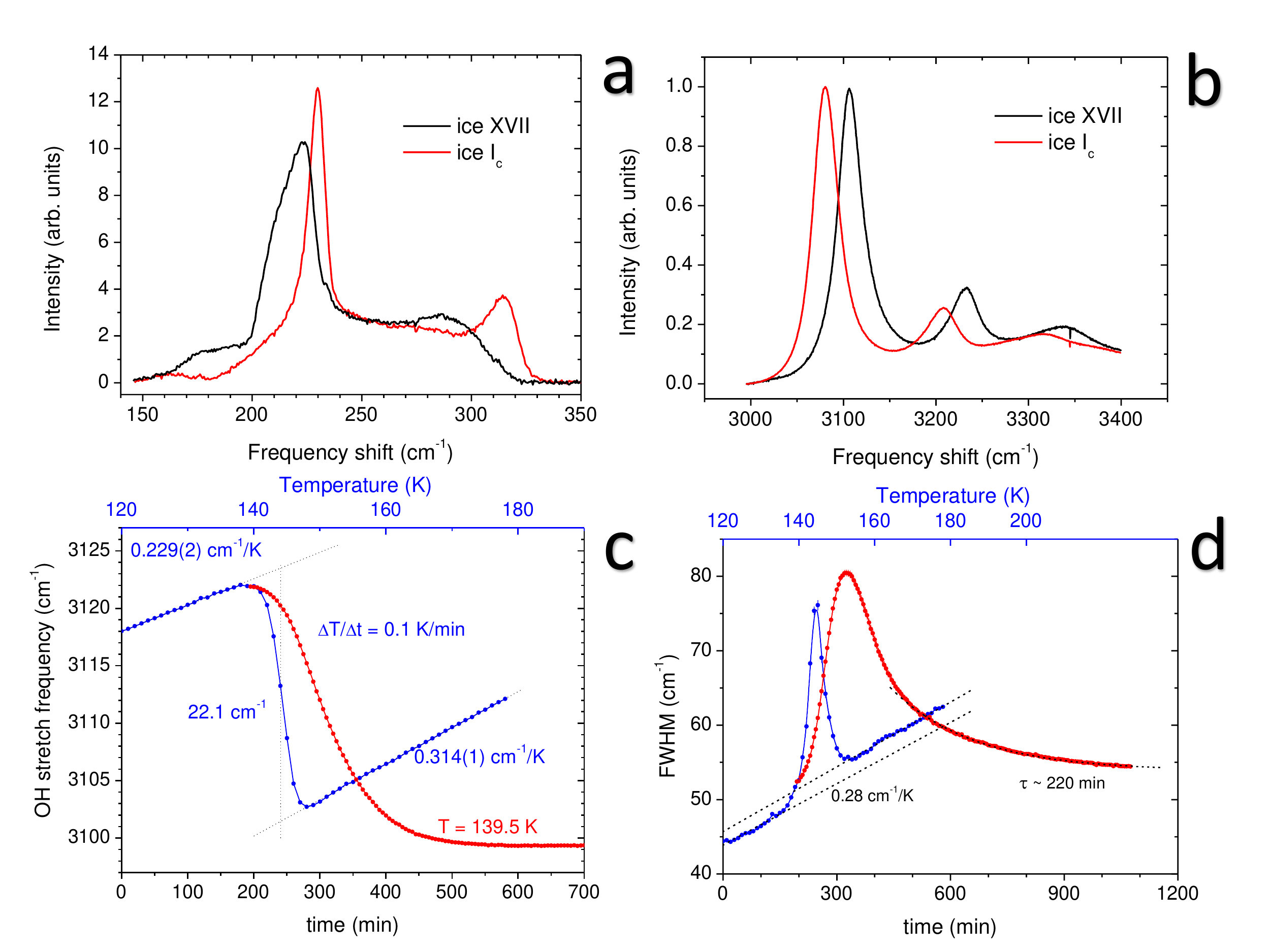}
 \caption{\footnotesize 
Raman spectra of ice XVII (black lines) and ice  Ic (red lines) in the lattice modes  ({\bf a}) and  OH-stretching  frequency region ({\bf b}), measured at 50 K.
({\bf c}): Frequency position of the OH stretching band (centre of the Lorentzian curve fitting the major band) during the transition ice XVII -- ice  Ic, while performing a 0.1 K/min temperature ramp (blue  line and dots), or as a function of time at constant temperature $T=139.5$ K (red line and dots).
({\bf d}): Width of the OH stretching band (from the Lorentzian fit) measured during the same thermal treatments as in ({\bf c}). 
   }
 \label{Raman01}
\end{figure}
The transition can be easily detected by Raman spectroscopy, which is also a valuable method to study the transition kinetics as a function of either temperature or time.
The Raman spectra of the two phases, ice XVII and ice Ic, present marked differences, both in the lattice modes (150-350 cm$^{-1}$) and OH stretching region (3000-3500  cm$^{-1}$).
In the first region (Fig. \ref{Raman01}({\bf a})) the differences concern both the position of the peaks and the shape of the whole band, while for the OH stretching mode  (Fig. \ref{Raman01}({\bf b})) the softening by more than 25 cm$^{-1}$ in ice Ic is immediately evident.
This finding is in agreement with the increase of density and, consequently, decrease of the OH$\cdot \cdot $O bond length.
The frequency shift of the OH band, reported as a function of time  (Fig. \ref{Raman01}({\bf c})), has been used to monitor the  ice XVII -- ice Ic transition, measured in two subsequent experiments.
In the first one, the temperature of the sample was slowly increased (0.1 K/min) from 120 to 178 K, thus allowing to determine the temperature of the structural transformation.
The  central frequency of the most intense peak of the OH band, fitted as a Lorentzian (see Method section) is reported as a blue line in Fig. \ref{Raman01}({\bf c}).
The transformation to ice Ic, marked by the large frequency decrease, (22.1 cm$^{-1}$) begins slightly below 139 K, and has it maximum rate at 144 K.
The slope of the OH frequency increase with temperature is slightly larger in ice Ic (0.314 cm$^{-1}$/K) than in ice XVII  (0.229 cm$^{-1}$/K).
The transformation appears to be complete in about 110 min, during which the temperature has increased by 11 K.
The kinetics of the transformation at one fixed temperature, namely $T= 139.5$ K (chosen just above the temperature at which a detectable spectral change  has been observed in the first experiment) is studied in the second measurement.
As expected, in this case the transformation is slower.
Data are represented by the red lines in Fig.~\ref{Raman01}({\bf c})({\bf d}). 
The time it takes to complete the transformation can be estimated at about 7 h.
The transformation process can be efficiently monitored by observing the width of the first peak versus time (Fig.~\ref{Raman01}({\bf d})).
The measured broadening during the transition is obviously due to the coexistence of the two phases during the structural rearrangement, which lasts, in the first case, about 200 min.
The kinetic series measured at $T=139.5$ K evidences a much slower change, with a long time tail in the roughly exponential decay of the line width, with a time constant $\tau \simeq 220$ min.
This is evidence of a very slow structural relaxation process.

\begin{figure}[h]
\centering
 \includegraphics[width= 17.0cm]{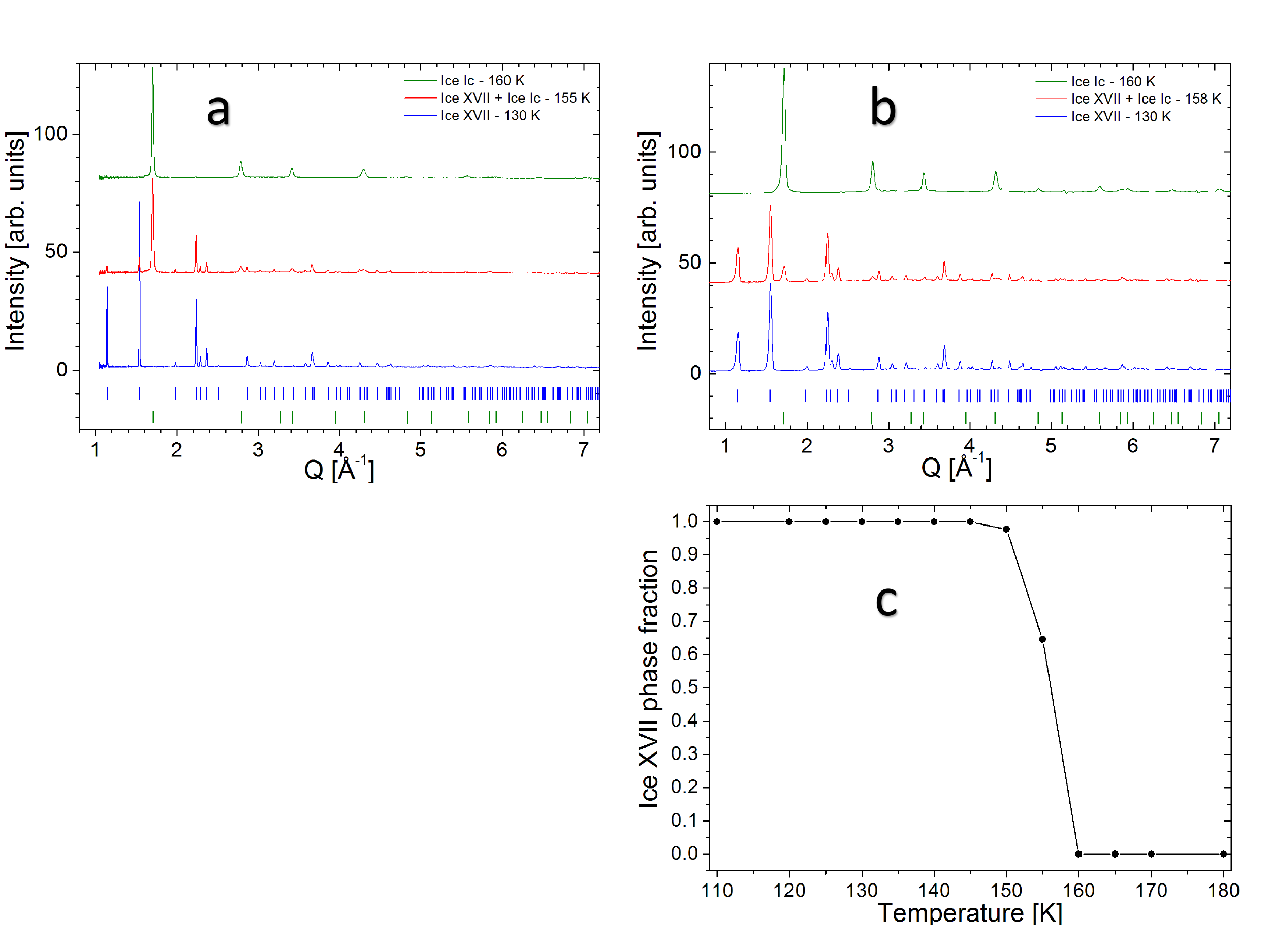}
 \caption{\footnotesize 
Diffraction patterns measured on HRPD ({\bf a}) and D20 ({\bf b}), plotted against $Q=2\pi/d$, showing the transition from ice XVII to ice  Ic.
Panel  ({\bf c}) shows the temperature dependence of the ice XVII phase fraction obtained by means of the Rietveld analysis of the HRPD data in the range 110 - 180 K.
 }
 \label{Figura_2}
\end{figure}
To determine the structure of our transformed sample, we have studied this same ice XVII -- ice Ic transition by neutron diffraction, during two different experiments performed, respectively, on HRPD at ISIS, RAL (UK) and on D20 at ILL (F). 
The experimental procedure was very similar in the two cases.
In both experiments the pure D$_2$O ice XVII sample, obtained after the completion of the annealing treatment (see Method section) was heated under dynamic vacuum, measuring the diffraction patterns at constant temperature intervals, (typically, 5 K, in the HRPD experiment or 10 K, in the D20 experiment) in the range 110--180 K.
In the HRPD experiment, the time interval needed to perform one measurement for a given thermodynamic point was about one hour, including about 30 minutes for an effective sample thermal equilibration, and 15 minutes for each of the two time-of-flight windows acquired.
For the D20 experiment, the thermodynamic history of the sample is quite similar to the previous one \cite{Catti19}.
In this case, short measurements were performed every 10 K in the range 110 -- 160 K, leaving 15 minutes for the stabilization at every temperature change.   
The main results of both experiments are summarized in Fig.~\ref{Figura_2}, where the most representative raw HRPD ({\bf a}) and D20 ({\bf b}) diffraction patterns are reported.

In both cases a clean ice XVII diffraction pattern ($P6_122$ symmetry group) is measured at low temperatures, below about 150 K.
On heating, a phase transition is observed between 150 and 160 K by the appearance of new diffraction peaks belonging to a different crystallographic phase; these were successfully indexed with the cubic symmetry $Fd\bar{3}m$.
The fraction of each phase during the transition has been determined by means of Rietiveld refinement of the experimental data, and is reported in Fig.~\ref{Figura_2}({\bf c}) as a function of temperature, for the data obtained on HRPD.
The refined parameters, obtained with the Rietveld method using either the GSAS \cite{Larson04}, or the the FullProf software packages\cite{Rodriguez-Carvajal93}, are listed in Table \ref{Tabella_fit}.
The apparent discrepancy of the transition temperature with respect to the Raman data is due to the faster heating rate applied in the neutron experiment.

\begin{figure}[H]
\centering
 \includegraphics[width= 13.0cm]{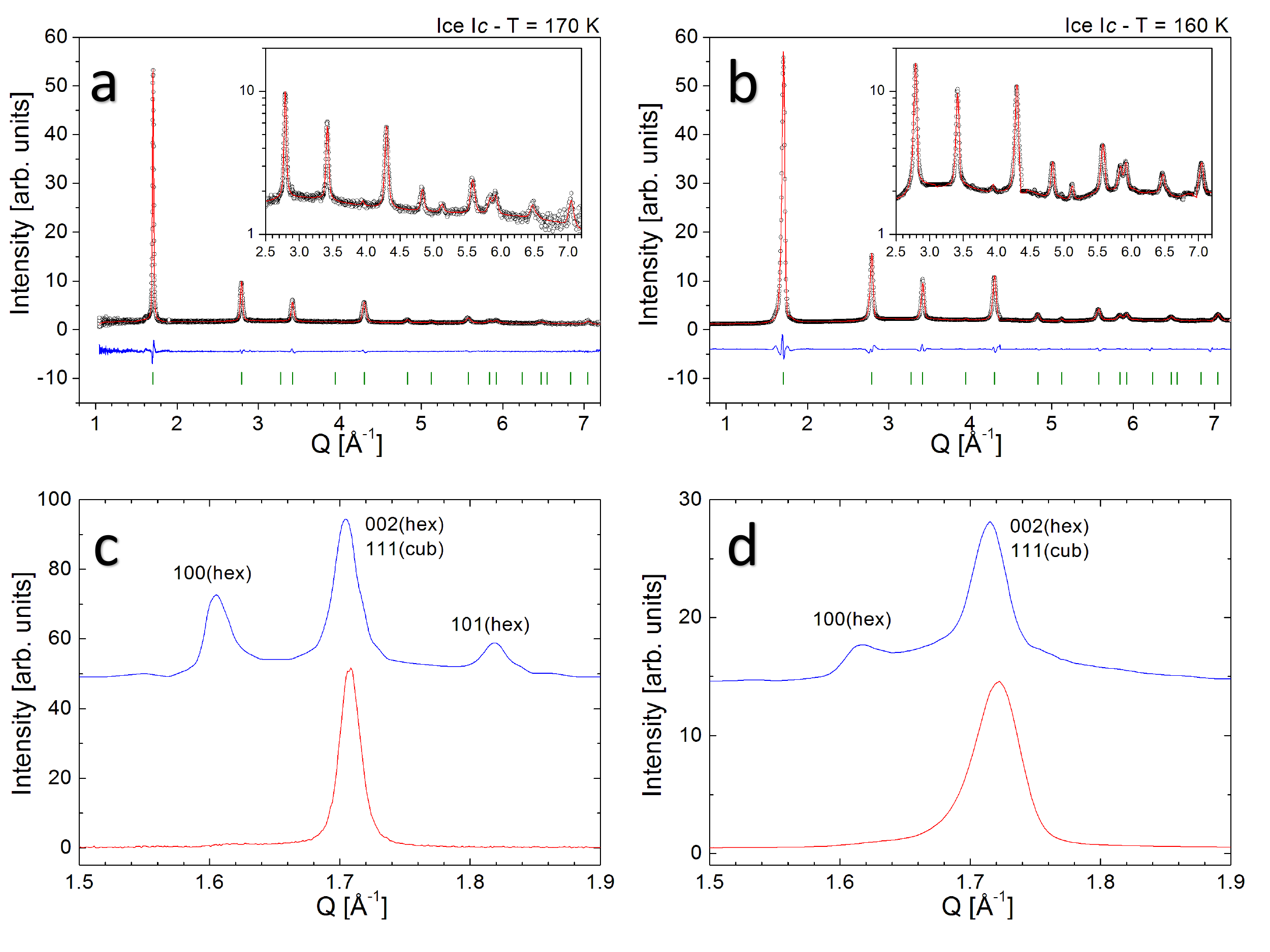}
 \caption
 {
 \footnotesize 
({\bf a, b}) Representative diffraction patterns (black circles) and Rietvield fits (red line) of the ice Ic data recorded at 170 K on HRPD ({\bf a}) and at 160 K on D20  ({\bf b}), both reported as a function of $Q$ for better comparison. 
To improve the visualization, a semilog plot with an enlargement of the high $Q$ zone is reported in the insets.
The residuals (blue line) and the positions of the ice  Ic reflections (green bars) are reported in each panel.
({\bf c}) HRPD neutron diffraction data of ice  Ic at 170 K (this study, red curve)  compared with a neutron diffraction measurement of ice Isd at 175 K, obtained from vapour deposition, with a cubicity of 78\% (Ref.~\protect\onlinecite{Kuhs12}, blue curve).
({\bf d})  D20 neutron diffraction data of ice  Ic at 160 K (this study, red curve)  compared with a synchrotron X-ray diffraction measurement of ice Isd at about 160 K, obtained from a recrystallization of ice II, with a cubicity of 73\%  (Ref.~\protect\onlinecite{Malkin15}, blue curve).
 }
 \label{Figura_3}
\end{figure}

The diffraction patterns recorded above 160 K, that present only the cubic ice reflections, indicate the completion of the transition from ice XVII to ice   Ic.
These are reported in Fig.~\ref{Figura_3}({\bf a}) and \ref{Figura_3}({\bf b}) for the experiments on HRPD and D20, respectively. 
The high degree of structural purity of ice  Ic, with no detectable sign of stacking disorder, is demonstrated by the absence of any hexagonal ice Ih reflections and diffuse scattering between the Bragg peaks in the D20 diffraction pattern and by the good quality of the fit, which is characterized by an agreement index $R_w$ equal to $ 4.1 \%$.
Also the sample of ice  Ic measured on the HRPD diffractometer appears highly pure, and is refined using the same cubic structure ($R_w = 4.4 \% $). 
Even the presence of any fraction of possible amorphous  ice phases in the sample can be totally excluded, from the observation that the continuous background intensity remains very small after the transition from ice XVII to ice Ic at about  150 K.

To our knowledge, a sample of ice  Ic with such a structural purity has never been obtained before.
Ice Isd, instead, is generally obtained, which presents a diffraction pattern with symmetry-forbidden reflections.
The region around the reflection that is indexed as $(111)$ in the cubic cell, indicated in the following as $(111)_{cub}$, is generally used to judge of the structural purity (the ``cubicity'') of the sample, as a sort of finger-print for the stacking disorder signatures.\cite{Kuhs87}
Three reflections of hexagonal ice Ih, indexed as   $(100)_{hex}$  $(002)_{hex}$  and $(101)_{hex}$ in the hexagonal cell, are present in this region, with the $(002)_{hex}$ reflection  coincident with $(111)_{cub}$ one. 
In the presence of stacking disorder, which can vary substantially as a function of the preparation method and history of the sample\cite{Kuhs12}, the shape of these reflections is modified and can be modelled by advanced statistical methods to define  its ``cubicity'' as some sort of proportion of cubic stacking sequences. \cite{Hansen08,Hansen15,Malkin15}.
So far, the highest cubicity obtained experimentally is $\approx 80 \%$. 
Numerous examples of diffraction patterns commonly obtained for ice Isd, synthesised by different routes, are reported in Fig.~3 of Ref.~\onlinecite{Malkin15}.
We compare in Fig.~\ref{Figura_3}({\bf c}),({\bf d}) the shape of the $(111)_{cub}$ reflection of ice  Ic obtained in our experiments with the diffraction pattern of two instances of ice Isd having the highest degree of cubicity obtained to date, namely 78\% \cite{Kuhs12} (Fig.~\ref{Figura_3}({\bf c})) and 73\% \cite{Malkin15} (( Fig.~\ref{Figura_3}({\bf d}))).

\begin{table}[H]
\centering
 \includegraphics[width= 16.0cm]{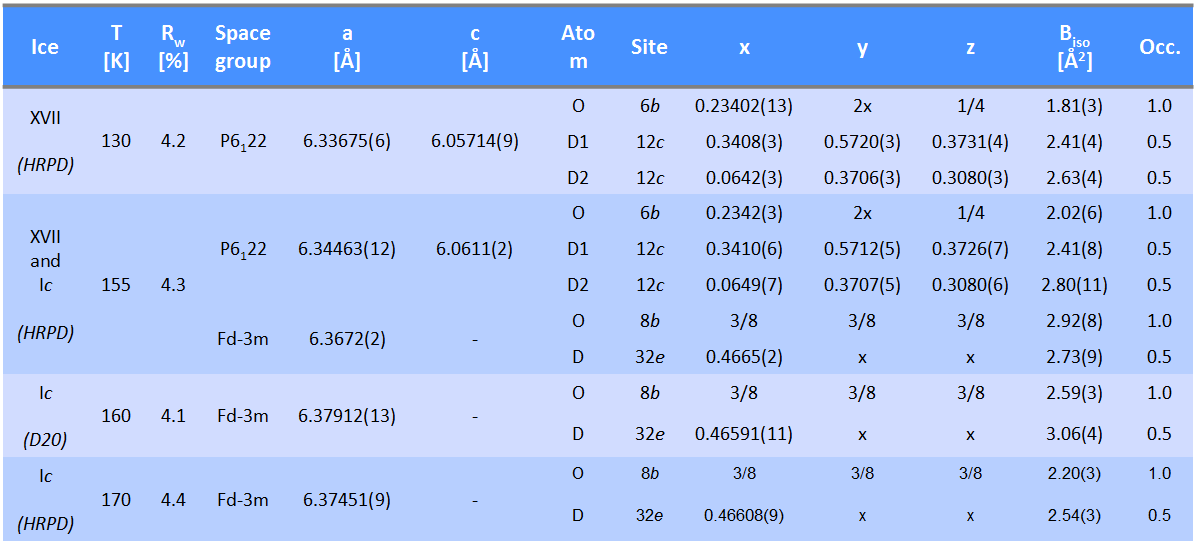}
 \caption{\footnotesize 
Lattice constants, atomic fractional coordinates and displacement parameters obtained by means of Rietveld refinement of the HRPD diffraction pattern (back scattering and equatorial banks, 30-130 ms t.o.f window) measured before (130 K), during (155 K) and after (170 K) the transition between Ice XVII and Ice  Ic, and of the D20 measurement at 160 K ($\lambda =1.542$ \AA).
}
 \label{Tabella_fit}
\end{table}

\begin{figure}[H]
\centering
 \includegraphics[width= 13.0cm]{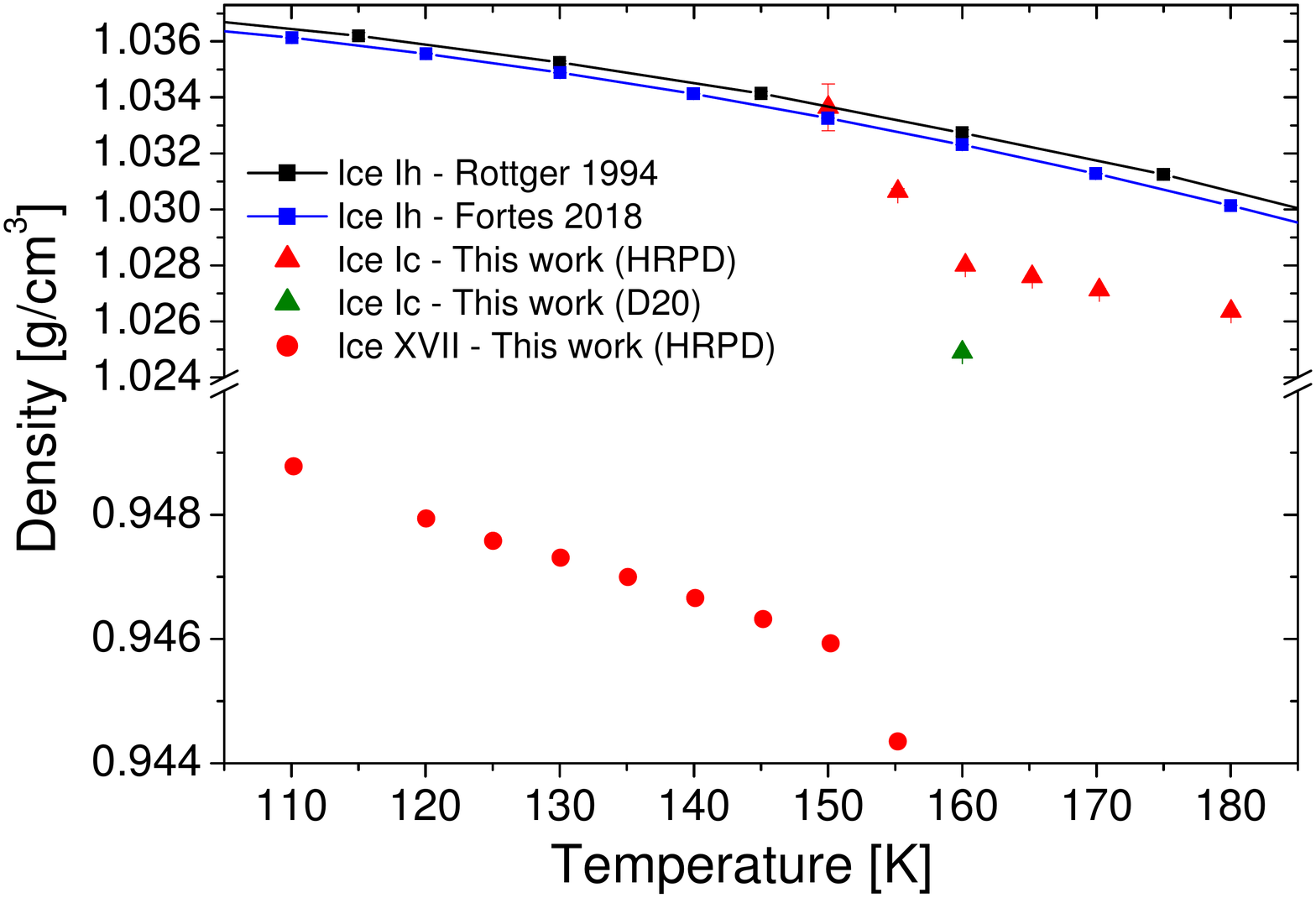}
 \caption{\footnotesize 
 Temperature dependence of the density of D$_2$O ice XVII (red dots) and ice Ic  (red triangles, HRPD data, green triangle, D20 datum). The values are obtained by means of Rietveld refinement of the diffraction data. For comparison, the density of D$_2$O ice Ih  as obtained from a synchrotron X-ray \cite{Rottger94} and neutron \cite{Fortes18} measurements are reported in the plot with black and blue lines, respectively.
}
 \label{Dens_vs_T}
\end{figure}

From the Rietveld refinement of the HRPD data we  have derived the behaviour of the unit cell parameters as a function of the temperature, during and after the transition from ice XVII to ice  Ic.
The density of ice Ic is reported in Fig.~\ref{Dens_vs_T}.
The data collected before and after the transition were refined with a single phase (i.e. space group $P6_122$ and $Fd\bar{3}m$ for ice XVII and  Ic, respectively), while the data collected during the transition (150 and 155 K) were refined with both phases.
The lattice parameters obtained at these two transition temperatures might be affected by some systematic uncertainty due to the non-homogeneous filling of the sample can by the two phases.
D$_2$O ice Ic displays a slightly lower density than that measured for ice Ih.
 The ice  Ih --  ice Ic positive density difference is larger than the statistical uncertainty derived from the Rietveld fit, but we cannot exclude a systematic uncertainty, which presumably may be of the same order of magnitude of the difference between the two determinations for ice Ih.      

In conclusion, this result here reported represents a milestone for the complete characterization of the ice polymorph I. 
The new route to obtain pure cubic ice in large quantities opens new perspectives for the clarification of fundamental questions in ice physics.
In the future, the specific heat of real ice Ic will be measured, which, together with the Ic --  Ih  transformation enthalpy, will enable the experimental determination of its free energy, and to resolve the issue of relative stability of the two polytypes  of ice present at room pressure.
This free energy difference is too low to be calculated by the computational methods available at present.\cite{Handa86,Engel15}
Precise structural measurements on good quality ice Ic samples will determine its thermal expansion, and clarify if there is a  region of negative expansivity  similar to hexagonal ice.
In addition, new studies will investigate if proton ordered phases of this same structure exist, in analogy with many other phases of ice\cite{Raza11,Salzmann19}, and if cubic ice undergoes at high pressure the same phase transformations as hexagonal ice, or it follows a different route.

\begin{acknowledgement}
Neutron beam time at ISIS and ILL is gratefully acknowledged, on the basis of the agreement of the CNR (Italy) with STFC (U.K.) and ILL (France) concerning collaboration in scientific research.
L. Ulivi and M. Celli acknowledge the PRIN project ZAPPING, No. 2015HK93L7, granted by the Italian MIUR (Ministero dell'Istruzione, dell'Universit\`a e della Ricerca) supporting their research in high-pressure materials science.
L. Ulivi, M. Celli and L. del Rosso acknowledge support from the Fondazione Cassa di Risparmio di Firenze under the contract ICEXVII.
ISIS Pressure and Furnaces section and the Electronics section were vital for setting up with gas-handling system and in-situ sample heating for the HRPD experiment. 
Technical support by Andrea Donati for the setting up of the high pressure autoclave for the synthesis of the samples is gratefully acknowledged. 
\end{acknowledgement}

\section{Author contribution}
This work is the result of a common effort to which all authors contributed, in particular:
L. del Rosso and M. Celli for the synthesis of samples;
M. Celli, L. del Rosso  and L. Ulivi, for the Raman experiment;
M. Catti, L. del Rosso, L. Ulivi, T. C. Hansen   for the experiment at ILL;
L. del Rosso, F. Grazzi, A. D. Fortes for the experiment at ISIS, RAL;
M. Celli and L. Ulivi for the Raman data analysis;
M. Catti, L. del Rosso, F. Grazzi, A. D. Fortes  for the diffraction data analysis;
L. Ulivi, L. del Rosso, M. Celli, for the writing of the manuscript.
All the authors read and corrected the manuscript.   

\section{Competing interests}
The authors declare no competing interests.


\section{Methods}

\subsection{Sample preparation}
Our starting material is the hydrogen filled ice (D$_2$O for the neutron experiment or H$_2$O for the Raman one) in the C$_0$-phase.
The high pressure synthesis procedure, with the recovery of the sample at room pressure, either for Raman spectroscopy or neutron diffraction measurements is similar to that described in Ref.~\onlinecite{del_Rosso16jpcc}. 
The sample is transferred from the Dewar into the experimental (Raman or neutron) cell at liquid nitrogen temperature in a dry nitrogen atmosphere to prevent the condensation on the sample surface of the water present in the air.
Once into the cryostat, the pristine H$_2 $ filled ice sample underwent the so called annealing treatment, i.e. heated up to 110 K under dynamic vacuum for about 1 hour, to remove all the guest molecules trapped inside the water structure, and thus obtaining the pure D$_2 $O ice XVII.
  
\subsection{Raman spectroscopy}
For the Raman experiment we have used the apparatus described in Ref.\onlinecite{Giannasi08}, equipped with a Raman cell with a single round window on the top surface. 
The sample handling and transfer into the cell are performed with the same precautions described above, and also the annealing procedure has been the same.
The temperature of the sample cell is controlled with a reproducibility of 0.1 K.
Few tens of millibar of Ar gas helps to reach thermal equilibrium between the powder sample and the cell.
The laser beam (Ar ion laser at $\lambda = 5145$ \AA) having a power on the sample of about 10 mW, is focussed onto a $80\  \mu$m spot on the powder sample, laying on a horizontal surface in the cell, in an almost back-scattering geometry.
The scattered light is collected in a vertical direction, and is focussed onto the entrance slit of a monochromator, equipped with a cooled CCD detector.
Spectra are collected in 30 or 10 min, for the lattice modes or the OH stretching region, respectively.
The data reported in Fig.~\ref{Raman01}({\bf c}) and \ref{Raman01}({\bf d}) are obtained by using a ramp on the temperature controller and by means  of an automatic spectra collection and analysis.
The frequency axis is calibrated by means of a Ne spectral lamp with an accuracy of 0.5 cm$^{-1}$, which is also the accuracy of the position and width of the OH stretching peak discussed in the text.

\subsection{Neutron diffraction}
The time-of-flight measurements performed with the High Resolution Powder Diffractometer at ISIS (Rutherford Appleton laboratory, U.K.) were based on three detector banks at high, middle and low resolution ($2\theta = 168.3 ^{\circ}$, $90 ^{\circ}$, $30 ^{\circ}$, $\Delta d/d \sim 6\cdot 10^{-4}$, $2\cdot 10^{-3}$, $2\cdot 10^{-2}$ respectively).
Data were collected in the 30-130 and 100-200 ms t.o.f. windows covering the 0.65-10.2 \AA\ and 2.15- 15.7 \AA\ $d$-spacing ranges, respectively, with all three banks.
Data are available at the ISIS database\cite{ISISdata}.
As sample container, we used a cylindrical vanadium cell provided with a capillary connected to a gas manifold equipped with a vacuum pump and various gas cylinders, that allows to set the pressure inside the cell in the range between 0 and 10 bar.
No empty cell measurement was needed since the Vanadium container gives a negligible coherent contribution to the scattering.
Low temperatures were achieved with a closed-cycle refrigerator mounted in HRPD's sample tank.
Like the sample holder, the windows in the tail of the CCR are made of vanadium foil and make only a minuscule contribution to the coherent scattering.

Raw data were time-focussed, normalised to the incident spectrum and corrected for instrumental efficiency using a V:Nb standard with the suite of diffraction routines in Mantid\cite{Mantid}.
 Data were analysed by the Rietveld refinement with the GSAS-I software package \citep{Larson04}, using data collected on the two detector banks at $2\theta = 168.3 ^{\circ}$ (30-130 ms t.o.f. window) and $90 ^{\circ}$ (30-130 and 100-200 ms t.o.f. windows).
The HRPD instrument calibration file, used to convert and fit the raw data from time-of-flight to d-spacing, was carefully obtained by measuring a NIST silicon standard (SRM640e) in the same cycle of our experiment.
A peak profile function 3, i.e. a convolution of back-to-back exponentials with a pseudo-Voigt function, was used for fitting the data: the profile parameter $\alpha$-0, $\beta$-0 and $\beta$-1, that model the instrumental broadening of the diffraction peaks, were carefully determined by measuring a CeO$_2 $ sample in the same cycle and were kept fixed during the refinement of the ice XVII experimental data; while the other gaussian and lorentzian profile parameters, that are mainly related to the peak broadening induced by the sample, were always refined.

A similar experiment was also performed by means of the high-flux high-resolution D20 diffractometer present at the Institute Laue Langevin (France). The technical details related to the experiment and data analysis are reported in Ref.~\onlinecite{Catti19}.
Data are stored in the appropriate ILL (France) database\cite{ILLdata}.

\end{document}